\documentclass[aps,prl, amsmath, showpacs, preprintnumbers,superscriptaddress, twocolumn,sort&compress,floatfix, amssymb]{revtex4}
\usepackage{graphicx}
\usepackage{rotating}
\usepackage{dcolumn}
\usepackage{bm}
\usepackage{color}
\usepackage{mathptmx, textcomp}
\usepackage[latin1]{inputenc}
\usepackage{braket}

\begin{document}

\author{F. Meinert}
\affiliation{Institut f\"ur Experimentalphysik und Zentrum f\"ur Quantenphysik, Universit\"at Innsbruck, 6020 Innsbruck, Austria}
\author{M. J. Mark}
\affiliation{Institut f\"ur Experimentalphysik und Zentrum f\"ur Quantenphysik, Universit\"at Innsbruck, 6020 Innsbruck, Austria}
\author{E. Kirilov}
\affiliation{Institut f\"ur Experimentalphysik und Zentrum f\"ur Quantenphysik, Universit\"at Innsbruck, 6020 Innsbruck, Austria}
\author{K. Lauber}
\affiliation{Institut f\"ur Experimentalphysik und Zentrum f\"ur Quantenphysik, Universit\"at Innsbruck, 6020 Innsbruck, Austria}
\author{P. Weinmann}
\affiliation{Institut f\"ur Experimentalphysik und Zentrum f\"ur Quantenphysik, Universit\"at Innsbruck, 6020 Innsbruck, Austria}
\author{M. Gr\"obner}
\affiliation{Institut f\"ur Experimentalphysik und Zentrum f\"ur Quantenphysik, Universit\"at Innsbruck, 6020 Innsbruck, Austria}
\author{H.-C. N\"agerl}
\affiliation{Institut f\"ur Experimentalphysik und Zentrum f\"ur Quantenphysik, Universit\"at Innsbruck, 6020 Innsbruck, Austria}

\title{Interaction-induced quantum phase revivals and evidence for the transition to the quantum chaotic regime in 1D atomic Bloch oscillations}

\date{\today}


\begin{abstract}
We study atomic Bloch oscillations in an ensemble of one-dimensional tilted superfluids in the Bose-Hubbard regime. For large values of the tilt, we observe interaction-induced coherent decay and matter-wave quantum phase revivals of the Bloch oscillating ensemble. We analyze the revival period dependence on interactions by means of a Feshbach resonance. When reducing the value of the tilt, we observe the disappearance of the quasi-periodic phase revival signature towards an irreversible decay of Bloch oscillations, indicating the transition from regular to quantum chaotic dynamics.
\end{abstract}

\maketitle

The response of a single particle in an ideal periodic potential when subject to an external force constitutes a paradigm in quantum mechanics. As first pointed out by Bloch and Zener the evolution of the wave function is oscillatory in time rather than linear, due to Bragg scattering of the matter wave on the lattice structure \cite{BLOCH28,ZENER34}. Yet, the observation of such Bloch oscillations in condensed-matter lattice systems is hindered by scattering of electrons with crystal defects \cite{FELDMANN92}, which causes rapid damping of the coherent dynamics, eventually allowing for electric conductivity \cite{ASHCROFTBOOK}.

Ensembles of ultracold atoms prepared in essentially dissipation-free optical lattices guarantee long enough coherence times \cite{Morsch2006} to serve as ideal systems for the observation of Bloch oscillations \cite{DAHAN96,ANDERSON98}. Furthermore, unprecedented precise control over atom-atom interactions in a Bose-Einstein condensate (BEC), via e.g.~Feshbach resonances \cite{CHIN10}, allows to engineer controlled coherent dephasing of the atomic matter wave and even to cancel interactions entirely. Bloch oscillations have been studied with BECs in quasi one-dimensional "tilted" lattice configurations created by a single retro-reflected laser beam with typically thousands of atoms per lattice site \cite{MORSCH01}. Here, comparatively weak interactions result in strong dephasing observed in a rapid broadening of the atomic wave packet in momentum space \cite{GUSTAVSSON08,FATTORI08}, and are understood by a modification of the wave function in terms of a mean field \cite{WITTHAUT05,GUSTAVSSON10}. In contrast, when the site occupancy approaches unity and interactions are strong, the microscopic details determine the quantum coherent time evolution, covered within the Bose-Hubbard (BH) model \cite{Jaksch1998}. One consequence is that one expects the emergence of a periodic coherent decay and revival of the Bloch oscillations entirely defined by the atom-atom interaction \cite{KOLOVSKY03}.

\begin{figure}
\includegraphics[width=0.3275\columnwidth]{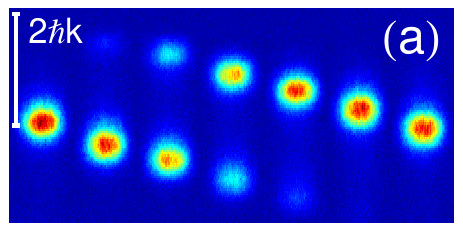}
\includegraphics[width=0.3275\columnwidth]{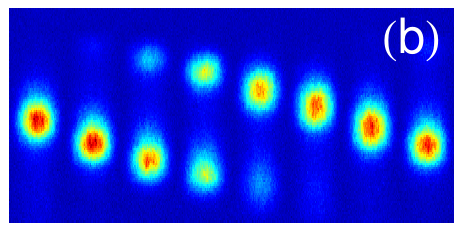}
\includegraphics[width=0.3275\columnwidth]{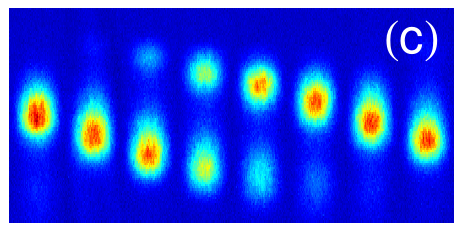}\\
\vspace{2mm}
\includegraphics[width=1\columnwidth]{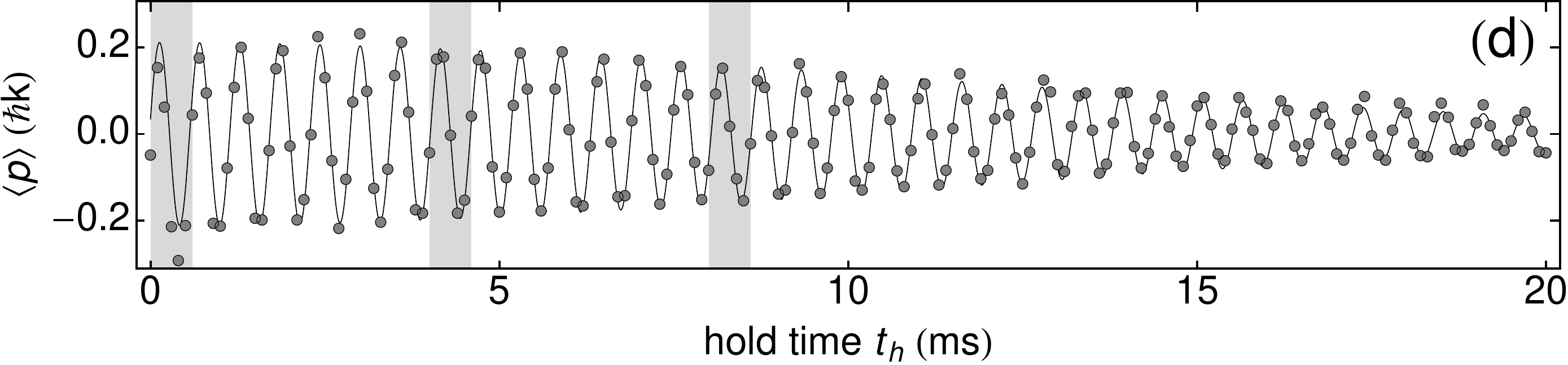}\\
\vspace{2mm}
\includegraphics[width=0.3275\columnwidth]{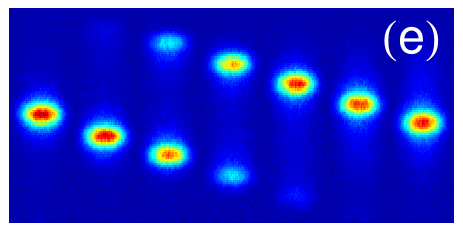}
\includegraphics[width=0.3275\columnwidth]{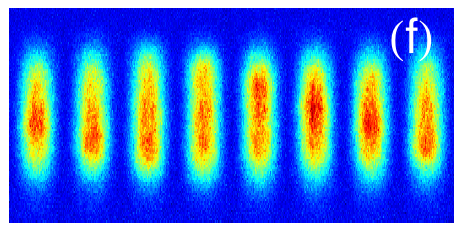}
\includegraphics[width=0.3275\columnwidth]{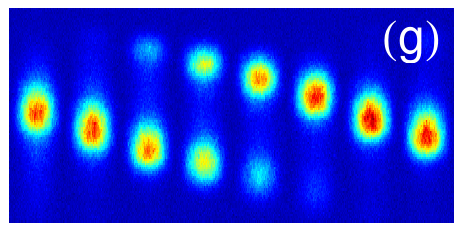}\\
\vspace{2mm}
\includegraphics[width=1\columnwidth]{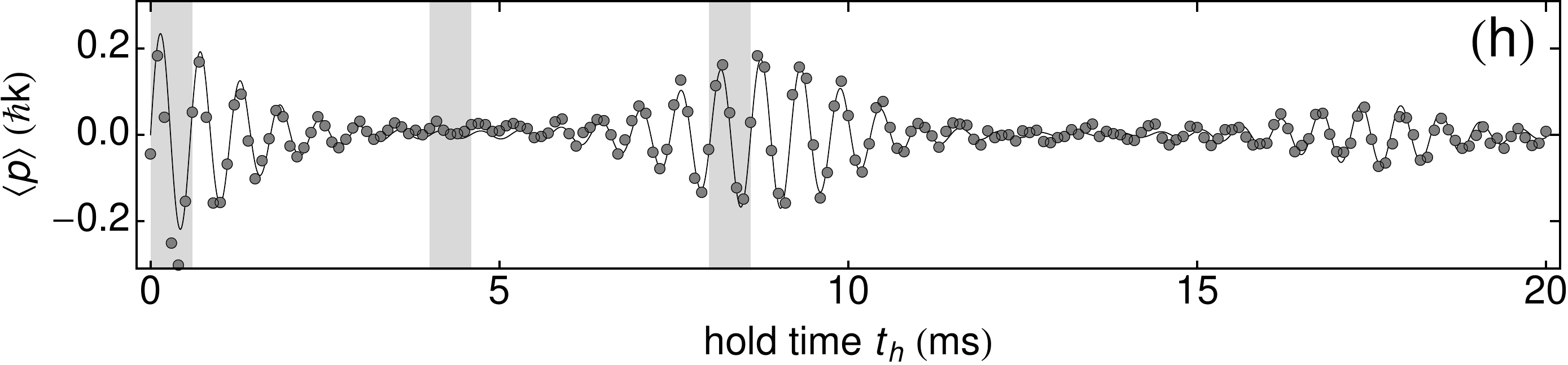}
\caption{\label{FIG1}Interaction-induced collapse and revival of atomic Bloch oscillations in a large ensemble of 1D tubes. The lattice depth along the tubes is $V_z = 7 E_R$ ($J=52.3\, \rm{Hz}$) and the transverse confinement is $V_{x,y}=30 E_R$. Time-of-flight absorption images of one Bloch cycle are shown after $t_h = 0,7,14 \, T_{\rm{B}}$ for zero interaction (a-c) and for $a_s = 21.4(1.5) \, a_0$ (e-g), respectively. The vertical bar in (a) indicates the extent of the first Brillouin zone. Full time evolution of the mean atomic momentum is shown for zero interaction (d) and for $a_s = 21.4(1.5) \, a_0$ (h), giving $U=102(8) \, \rm{Hz}$. Solid lines are fits to the data using the analytic model function (see text). The shaded areas indicate the data points shown in the time-of-flight pictures.}
\end{figure}

\begin{figure}
\includegraphics[width=0.48\columnwidth]{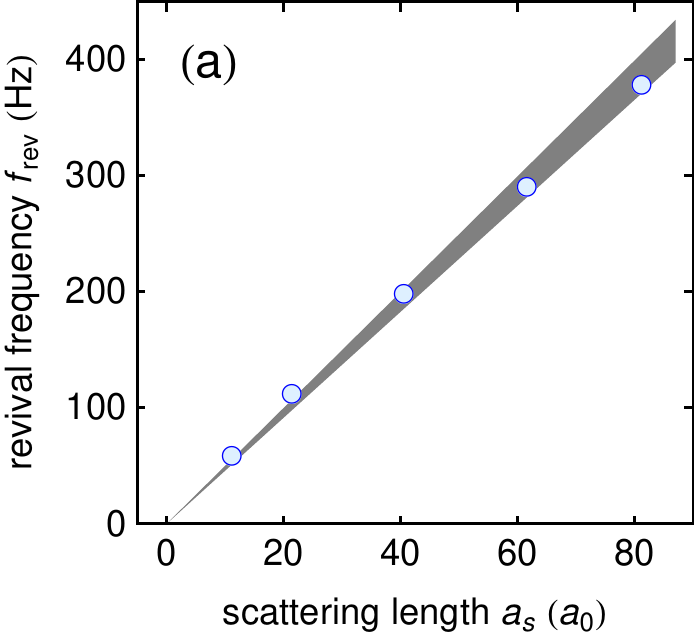}
\hspace{2mm}
\includegraphics[width=0.48\columnwidth]{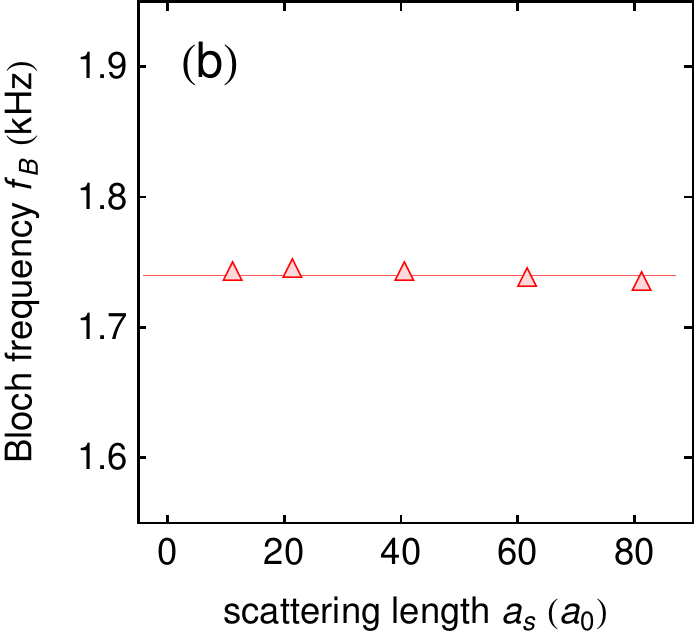}\\
\vspace{2mm}
\includegraphics[width=1\columnwidth]{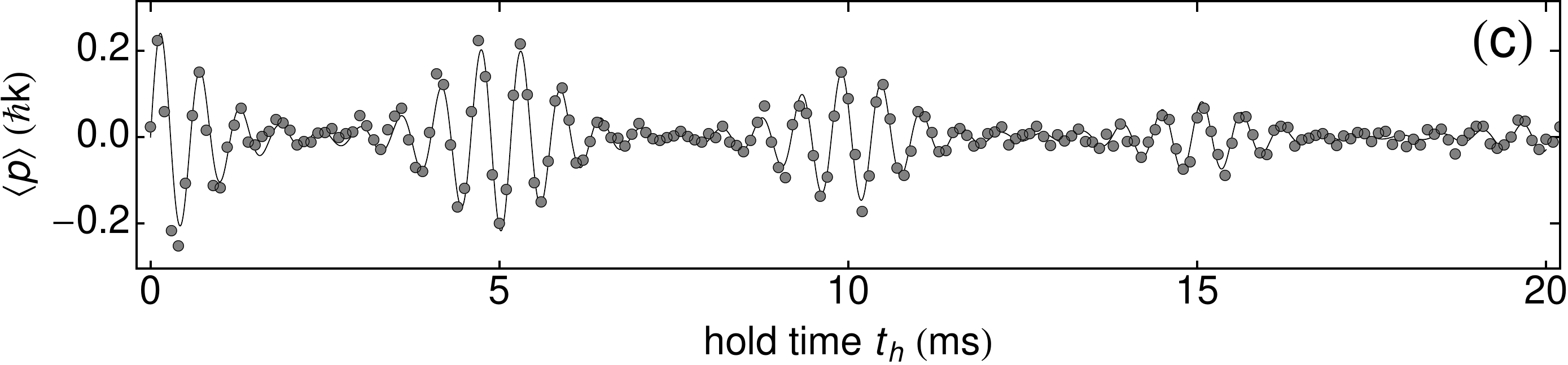}\\
\vspace{2mm}
\includegraphics[width=1\columnwidth]{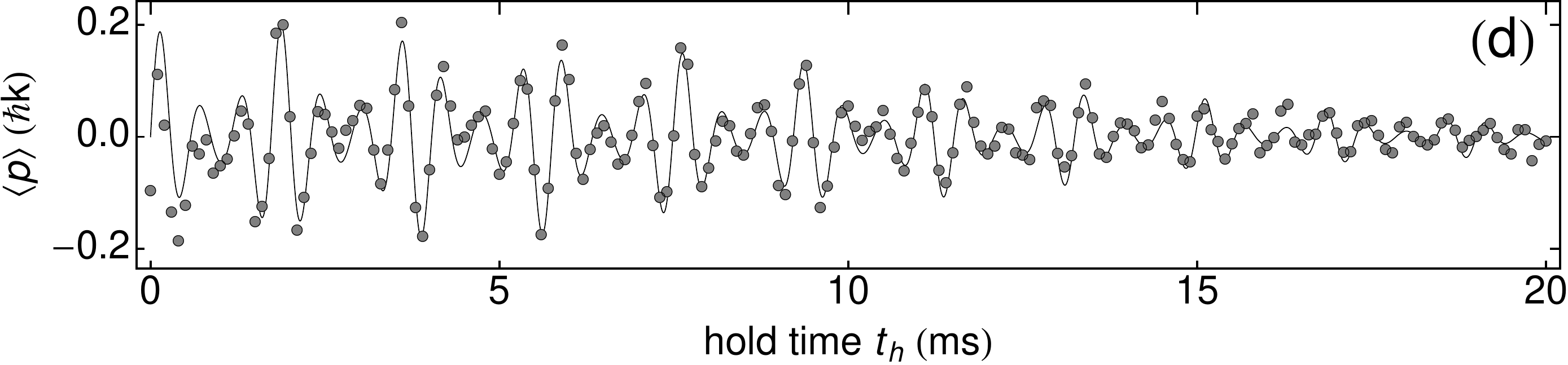}
\caption{\label{FIG2}Dependence of the revival frequency $f_{\rm{rev}}$ on the on-site interaction energy $U$. Extracted $f_{\rm{rev}}$ (a) and Bloch frequency $f_B$ (b) from model fits of Eq.~(\ref{EQ2}) to time traces $\langle p \rangle (t_h)$ as shown in Fig.~\ref{FIG1} as a function of $a_s$. Error bars are smaller than the data points. The shaded gray area in (a) depicts the calculated $U$ taking into account a $5\%$ error on the lattice depth. The solid line in (b) denotes a constant fit to the data. Mean atomic momentum as a function of hold time for $a_s = 40.6(1.5) \, a_0$ (c), giving $U=194(10) \, \rm{Hz}$, and $a_s = 111.5(1.5) \, a_0$ (d), giving a comparatively large $U=533(15) \, \rm{Hz}$. Solid lines are fits to the data based on Eq.~(\ref{EQ2}).}
\end{figure}

In this letter, we study atomic Bloch oscillations in the one-particle-per-site regime realized with a large ensemble of 1D superfluid Bose gases trapped in an array of "quantum tubes" and subject to a tilted optical lattice. In particular, we observe regular interaction-induced dephasing and revivals of the quantum matter-wave field for strongly tilted lattices. In contrast, for sufficiently small values of the tilt, an irreversible decoherence of the wave function in the presence of interactions is seen, giving a strong indication for the onset of quantum chaos \cite{KOLOVSKY03b}.

At ultralow temperatures, much below the lattice band gap, our system is well described by the one-dimensional BH Hamiltonian augmented by a tilt \cite{Jaksch1998,KOLOVSKY03}
\begin{equation}
\hat{H} = -J \sum\limits_{\langle i,j \rangle } \hat{a}_i^\dagger \hat{a}_j + \sum\limits_{i} \frac{U}{2} \hat{n}_i\left(\hat{n}_{i}-1\right) +E \sum\limits_i i \hat{n}_{i} \, .
\label{EQ1}
\end{equation}
As usual, $\hat{a}_i^\dagger$ ($\hat{a}_i$) are the bosonic creation (annihilation) operators at the $i$th lattice site, $\hat{n}_i = \hat{a}_i^\dagger \hat{a}_i$ are the number operators, $J$ is the tunnel matrix element, and $U$ is the on-site interaction energy. The linear energy shift from site to site is denoted by $E$. Let us first discuss the case of a strong force $E \gg J$. For vanishing interaction energy ($U=0$) the energy spectrum is the famous equidistant Wannier-Stark ladder, which gives rise to a single frequency, $f_B=E$ \cite{units}, contributing to the dynamics of any arbitrary initial wave packet \cite{GLUECK02}. This is the origin of Bloch oscillations for a wave packet in momentum space that is periodically driven across the first Brillouin zone and Bragg reflected at the zone edge with the Bloch frequency $f_B$ \cite{KOLOVSKY03a}. An intermediate interaction energy $U\approx J$ causes splitting of the degenerate energy levels of the Wannier-Stark ladder into a regular pattern of energy bands and results in quasi-periodic coherent decay and revival of the Bloch oscillations with a new fundamental frequency $f_{\rm{rev}}=U$ \cite{KOLOVSKY03,KOLOVSKY03b}. This can be understood with the previous assumption of a strong tilt ($E \gg J$) for which the eigenstates of a single atom in the lattice coincide with the localized Wannier states. The Stark localization of the wave function together with the discreteness of the site occupation number leads to an evolution of the mean atomic momentum
\begin{equation}
\langle p \rangle (t) \propto \exp \left( -2 n (1-\cos(2 \pi f_{\rm{rev}} t)) \right) \sin \left( 2\pi f_B t \right) \, ,
\label{EQ2}
\end{equation}
where $n$ denotes the mean occupation number per lattice site \cite{KOLOVSKY03,init}. Note the close analogy of the dynamical evolution to the experiment of Ref.~\cite{GREINER02}. For the sake of completeness, we stress that for sufficiently strong interaction energy ($U \gg J$) and sufficiently small tilt  ($E \ll U$) the ground state is a Mott insulator for one-atom commensurate filling, which shows resonant tunneling dynamics when subject to a sudden tilt \cite{MEINERT13}.

Our experiment starts with a BEC of typically $8 \times 10^4$ Cs atoms prepared in the internal hyperfine ground state $|F=3, m_F =3 \rangle$ and trapped in a crossed-beam optical dipole trap. The sample is levitated against gravity by a magnetic field gradient of $|\nabla B| \approx 31.1$~G/cm. Trapping and cooling procedures are described in Ref.'s.~\cite{Weber2002,Kraemer2004}. The BEC is loaded adiabatically into an optical lattice of three mutually orthogonal retro-reflected laser beams at a wavelength of $\lambda_{\rm{l}} = 2\pi / k = 1064.5$ nm within $500$ ms. During the lattice loading the scattering length is $a_s = 115 \, a_0$. At the end of the ramp, the final lattice depth is $V_{x,y}= 30 E_R$ in the horizontal and $V_z = 7 E_R$ ($J=52.3$ Hz) in the vertical direction, where $E_R=1.325$ kHz is the photon recoil energy. This creates an array of about 2000 vertically oriented 1D Bose-Hubbard systems ("tubes") at near unity filling that are decoupled over the timescale of the experiment. The residual harmonic confinement along the vertical $z$ direction is measured to $\nu_z=16.0(0.1)$ Hz. We now ramp $a_s$ slowly (with $\approx 1.5$~$a_0$/ms) to values of typically $0$ to $90 \, a_0$ by means of a Feshbach resonance \cite{Mark2011}, and thereby prepare the 1D systems near the many-body ground state for an on-site interaction energy $U$ of typically $0$ to $400$~Hz. This constitutes the initial condition for the observation of Bloch oscillations. Bloch oscillations are then initiated by quickly applying a gravity induced tilt $E = 1740(4)$~Hz through a reduction of the magnetic levitation field, giving a Bloch period $T_B \equiv 1/f_B = 575(1) \, \mu \rm{s}$. After a variable hold time $t_h$ we switch off the lattice beams within $300 \, \mu$s, remove the tilt, and allow the sample for a free levitated expansion of $30$ ms to measure the atomic momentum distribution by standard time-of-flight absorption imaging. During expansion, $a_s$ is set to zero to avoid any additional broadening due to interactions.

First, we study a non-interacting sample by setting $a_s = 0 \, a_0$ to quantify the effect of the residual harmonic trapping potential. Time-of-flight absorption images spanning one Bloch oscillation cycle are shown in Figs.~\ref{FIG1}(a)-(c) after $t_h = 0$, $7$, and $14$ $T_B$. Note that the aspect ratio has been adapted for better visualization to compensate the faster expansion transversal to the orientation of the tubes. The typical linear motion of the atomic wave packet through the first Brillouin zone together with matter-wave scattering at the zone edge shows a slow dephasing due to the harmonic confinement. The mean atomic momentum $\langle p \rangle$ extracted from time-of-flight images is depicted in Fig.~\ref{FIG1}(d) as a function of $t_h$. We model the dephasing effect of the harmonic trapping potential by a local variation of the Bloch frequency over the extent of the initial wave packet, giving $\langle p \rangle(t) \propto \sum_{i} \exp(-i^2/(2\gamma^2)) \sin\left((2\pi f_B + \nu i /  \hbar)t \right)$. Here, $\nu = m (2 \pi \nu_z)^2 d^2$, with $m$ the mass of the Cs atom, and $d=\lambda_{\rm{l}}/2$ the lattice spacing. The wave packet is approximated by a Gaussian envelope of half width $\gamma$. Replacing the sum by an integral yields $\langle p \rangle(t)\propto \exp(-t^2/(2 \tau_0^2)) \sin(2\pi f_B t)$, with $\tau_0=\hbar/\gamma \nu$  \cite{PONOMAREV06,MARK11}. A fit of the model to our data delivers $\tau_0 = 11.0(0.2)$ ms, from which we deduce a mean extension of the matter wave over $\approx 30$ lattice sites, in agreement with what we expect from our loading procedure.

\begin{figure}
\includegraphics[width=1\columnwidth]{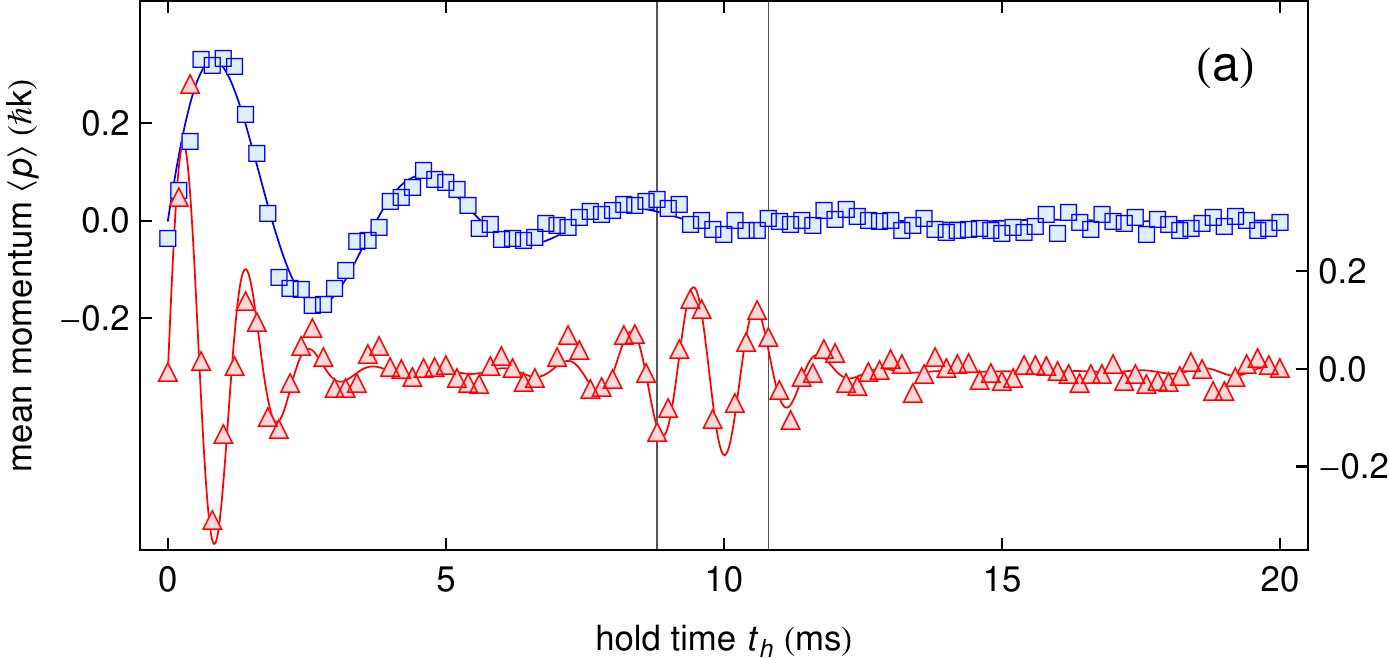}\\
\vspace{2mm}
\includegraphics[width=0.48\columnwidth]{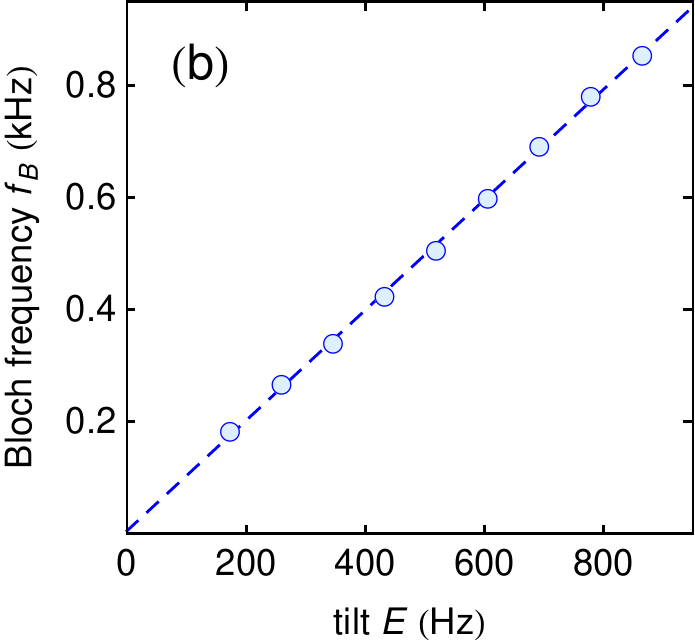}
\hspace{2mm}
\includegraphics[width=0.48\columnwidth]{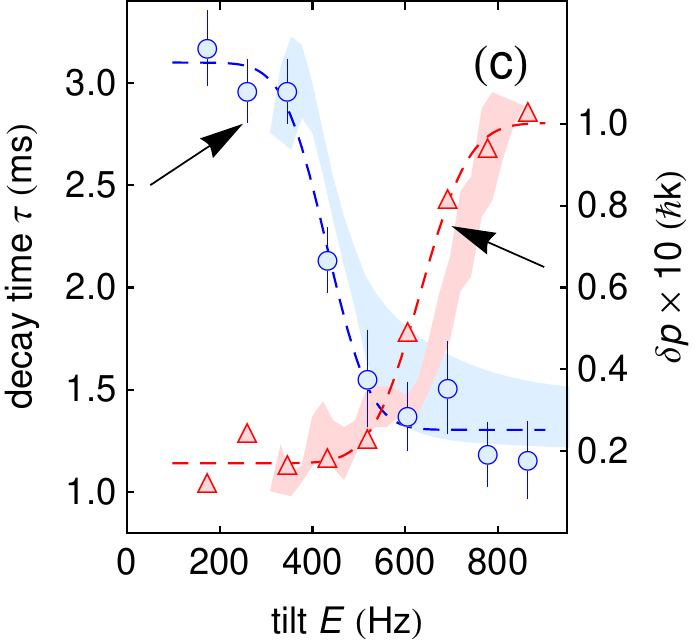}
\caption{\label{FIG3}Transition from regular to chaotic dynamics when varying the tilt $E$. (a) Mean atomic momentum as a function of hold time for $E=855(15) \, \rm{Hz}$ (triangles) and $E=266(5) \, \rm{Hz}$ (squares) at $a_s = 26.9(1.5) \, a_0$. Here, the lattice depth along the tubes is $V_z = 4 E_R$ ($J=114.2\, \rm{Hz}$) and the transverse confinement is $V_{x,y}=30 E_R$, giving $U=106(8) \, \rm{Hz}$. Solid lines show fits to the data based on the analytic model function, Eq.~(\ref{EQ2}), in the strong-forcing limit and an exponentially damped sinusoidal in the chaotic regime. The two datasets  are offset for clarity. Bloch frequency $f_B$ (b) and exponential decay time $\tau$ (c) as a function of $E$ (circles) extracted from exponentially damped sinusoidal fits to the initial decay in datasets as depicted in (a). Triangles in (c) depict the width of the revival $\delta p$ as a function of $E$. The dashed line in (b) is a linear fit to the data. The dashed lines in (c) are error-function fits to the data to guide the eye. The shaded areas indicate the prediction from numerical simulations with $1 \leq n \leq 1.4$ \cite{supmat}.}
\end{figure}

To demonstrate the effect of atom-atom interactions on the dynamics, we repeat the above measurement now with the sample prepared at $a_s = 21.4(1.5) \, a_0$, corresponding to $U=102(8) \, \rm{Hz}$. Analogously to the previous measurement, time-of-flight images of full Bloch oscillation cycles after $t_h = 0$, $7$, and $14$ $T_B$ are shown in Figs.~\ref{FIG1}(e)-(g). In contrast to the non-interacting sample, we observe a rapid initial decay of the Bloch dynamics, resulting in a spreading of the atomic cloud over the entire Brillouin zone, see Fig.~\ref{FIG1}(f). Following the dynamical evolution we find a near perfect recovering of the Bloch oscillations as evident from Fig.~\ref{FIG1}(g). This arises from the strong coherent dephasing of the initial atomic wave packet in the presence of atom-atom interactions discussed above, leading to a subsequent high contrast matter-wave phase revival. The mean atomic momentum $\langle p \rangle$ as a function of $t_h$ is plotted in Fig.~\ref{FIG1}(h) and fit by the model function Eq.~(\ref{EQ2}) including the overall decaying envelope discussed above to account for the residual harmonic trap. We do such measurements at different values for $a_s$. The values for $f_{\rm{rev}}$ and $f_B$ extracted from the two-mode fit function are depicted in Fig.~\ref{FIG2}(a) and (b), respectively. While the Bloch frequency $f_B$ is not affected by interactions, the revival frequency $f_{\rm{rev}}$ increases linearly with $a_s$ and is in good agreement with the prediction for $U$ calculated from lowest-band Wannier functions \cite{Jaksch1998}. Two additional datasets showing the evolution of the atomic sample in the tilted lattice are shown in Fig.~\ref{FIG2}(c) and (d) for intermediate and comparatively strong interaction energy, respectively, in order to demonstrate the experimental robustness of the quantum phase revival. While we find clear separation between the timescales given by $f_B$ and $f_{\rm{rev}}$ and observe four distinct decay and revival periods of the matter-wave packet in Fig.~\ref{FIG2}(c), Bloch and revival period start to mix when $U$ is increased to $\approx E/3$ in Fig.~\ref{FIG2}(d). Interestingly, the model function, Eq.~(\ref{EQ2}), still provides a surprisingly precise fit to the data. Note that our technique allows the direct measurement of the Bose-Hubbard interaction parameter $U$ in the superfluid regime down to very small values, limited only by the residual external confinement.

So far, we have restricted the discussion to large values of the tilt. In the remainder of this paper, we study Bloch oscillations in the regime when all energy scales in the BH Hamiltonian are of comparable magnitude, $E \approx J \approx U$. Consequently, it is impractical to assign a meaningful set of quantum numbers to the energy levels in a perturbative approach. Instead, the energy spectrum emerges densely packed and requires a statistical analysis, revealing the onset of quantum chaos in a Wigner-Dyson distribution of the energy level spacings for sufficiently small $E$ \cite{KOLOVSKY03b}. The transition from the regular to chaotic regime is predicted to become manifest in a rapid irreversible decoherence of Bloch oscillations within a few oscillation cycles. To probe this regime, we prepare the sample in a more shallow lattice $V_z = 4 E_R$ ($J=114.2$~Hz) at a fixed value of the scattering length $a_s = 26.9(1.5) \, a_0$, corresponding to $U=106(8) \, \rm{Hz}$, and now vary the value of $E$. Two example datasets are shown in Fig.~\ref{FIG3}(a) taken at $E=855(15)$~Hz and $266(5)$~Hz, respectively, which show very different qualitative behavior. For the strongly forced lattice we clearly identify the regular decay and revival dynamics described above. In contrast, for smaller $E$ the revival, expected to appear at $t_h=1/U$, is missing. Instead, we observe a single, rapid irreversible decay of Bloch oscillations. This observation is quantified twofold. First, we fit the initial decay in our data to a decaying sinusoid with an envelope $\propto \exp(-t / \tau)$. The extracted Bloch oscillation frequency $f_B$ and the exponential decay time $\tau$ as a function of $E$ are plotted in Fig.~\ref{FIG3}(b) and (c) (circles). Second, we extract the width of the revival $\delta p$ in momentum space \cite{width} and show it in Fig.~\ref{FIG3}(c) (triangles). As expected, $f_B$ is directly given by $E$. Further, $\tau$ is found constant for $E \gtrsim 600$~Hz, quickly increases, and finally saturates for $E \lesssim 400$~Hz. The revival signal $\delta p$ exhibits an opposite behavior and decreases with decreasing tilt.

\begin{figure}
\includegraphics[width=0.48\columnwidth]{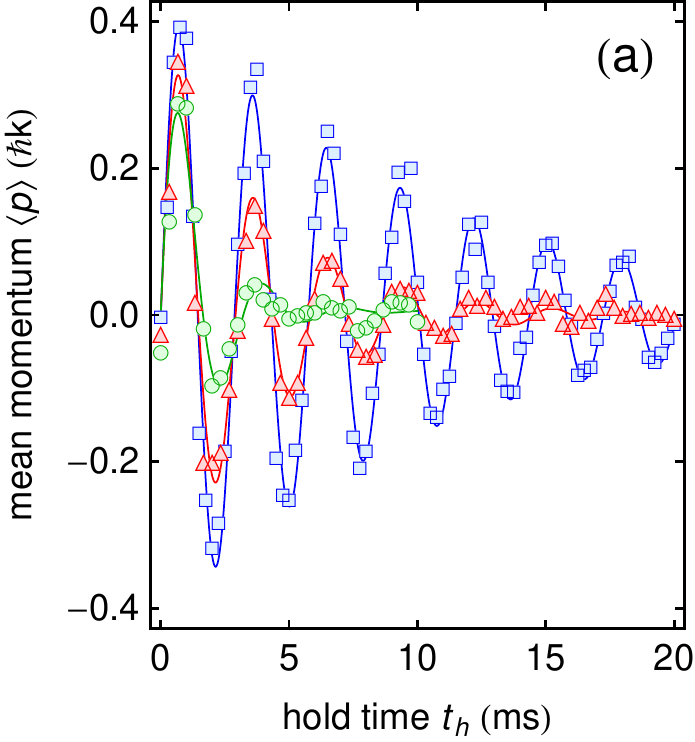}
\hspace{2mm}
\includegraphics[width=0.48\columnwidth]{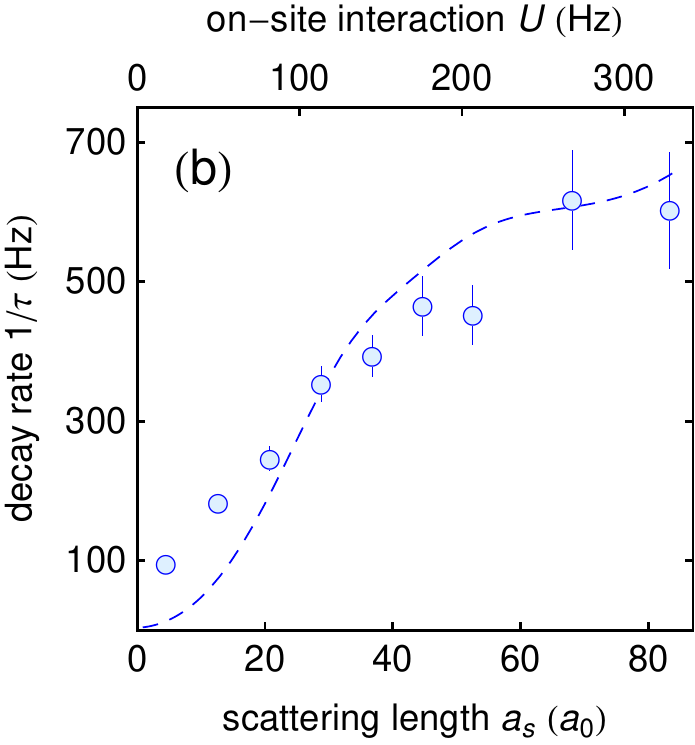}
\caption{\label{FIG4}Dependence of the irreversible decay of Bloch oscillations on atom-atom interaction in the chaotic regime. (a) Mean atomic momentum as a function of hold time for $a_s = 4.5(1.5) \, a_0$ (squares), $a_s = 20.8(1.5) \, a_0$ (triangles), and $a_s = 83.4(1.5) \, a_0$ (circles) at $E=346(10) \, \rm{Hz}$. Here, the lattice depth along the tubes is $V_z = 4 E_R$ ($J=114.2\, \rm{Hz}$) and the transverse confinement is $V_{x,y}=30 E_R$. Solid lines are exponentially damped sinusoidal fits to the data. (b) Exponential decay rate $1/\tau$ as a function of $a_s$ extracted from datasets as shown in (a). The dashed line shows a prediction from numerical simulations for a homogeneous system without a trap.}
\end{figure}

We interpret our data as a strong indication for the transition from the regular to the quantum chaotic regime. For large $E/J \gtrsim 6$, we find a coherent dephasing and revival of Bloch oscillations with a characteristic dephasing timescale that is solely given by $U$ and not affected by the strength of the tilt, as expected from Eq.~(\ref{EQ2}). With decreasing $E$, the revival continuously disappears accompanied by a significant change of $\tau$, indicating the onset of a regime where the system dynamics decay irreversibly for $E/J \lesssim 3$. A spectral analysis of the tilted BH model for our system parameters ($n \gtrsim 1$) reveals the onset of Wigner-Dyson statistics in agreement with the observed crossover in $\tau$ \cite{supmat}. In addition we compare our data in Fig.~\ref{FIG3}(c) to results obtained from numerical simulations of Bloch oscillations within the BH model, indicated by the shaded areas. We emphasize again that the irreversible decay is due to interaction-induced decoherence in the system arising from the multitude of avoided crossings in the chaotic level structure of the many-body energy spectrum and ought to be contrasted from a coherent dephasing as observed for $E \gg J$ \cite{KOLOVSKY03b}. In this sense, the atomic ensemble itself acts as the bath responsible for decoherence of the quantum many-body system \cite{RIGOL08}.

Finally, we report on a last set of measurements to emphasize the role of interactions on the irreversible decay discussed above. We study Bloch oscillations for a fixed tilt $E=346(10)$ Hz at $V_z = 4 E_R$ ($J=114.2$~Hz) and vary $U$. Three example datasets are shown in Fig.~\ref{FIG4}(a). We identify an overall irreversible decay indicative for the chaotic regime. Moreover, the decay rate strongly depends on the interaction strength and decreases with decreasing $U$. This is expected from the non-interacting limit $U \rightarrow 0$, for which the system turns regular again. Note that the observation of small decay rates $\lesssim 70$~Hz is currently limited by the overall dephasing due to the presence of the harmonic trap discussed above. We extract the decay rate $1/\tau$ as before from an exponentially damped sinusoid fit to the data and plot it as a function of $U$ in Fig.~\ref{FIG4}(b). Our data reveal a saturating monotonic increase of $1/\tau$ with $U$. Scaling of the decoherence rate with $U$ is expected to change from a quadratic (regular regime) to a square-root (chaotic regime) dependence {\cite{privat}, indicated by the dashed line in Fig.~\ref{FIG4}(b) \cite{supmat}. However, a precise experimental confirmation requires compensation of the harmonic trap to allow for longer observation times of the Bloch oscillations and will be the issue of a forthcoming experiment.

In conclusion, we have studied Bloch oscillations in the context of a strongly interacting many-body system properly described within the Bose-Hubbard model. In this regime with on average only  a few particles per lattice site, the "granularity" of matter in combination with strong atom-atom repulsion causes coherent quasi-periodic decay followed by a high contrast quantum phase revival of the Bloch oscillating matter-wave field. The revival period is entirely determined by the interaction strength and thus provides a direct, precise measure for the on-site interaction energy in the superfluid regime, in contrast to a related technique to measure $U$ in a Mott insulator \cite{WILL10}. For practical applications we point out that the phase revivals effectively extend the observation time of Bloch oscillations even in the presence of interactions with potential prospects to e.g. precision force measurements \cite{CARUSOTTO05, Mahmud2013}. Further, we have investigated the Bloch dynamics of the interacting atomic ensemble as a function of the applied tilt and found clear evidence for the transition from regular to quantum chaotic dynamics. Our results may open the experimental study of quantum chaos in such systems including its implication on the decoherence and thermalization of interacting 1D quantum many-body systems \cite{RIGOL08,KINOSHITA06}. Moreover, quantitative studies on the parameter dependence of the transition from the regular to the quantum chaotic regime are of interest \cite{ECKSTEIN11}.

We are indebted to R. Grimm for generous support, and thank M. Hiller and A. Buchleitner for fruitful discussions. We gratefully acknowledge funding by the European Research Council (ERC) under Project No. 278417.

\bibliographystyle{apsrev}

\newpage
\clearpage

\section{Supplementary Material: Interaction-induced quantum phase revivals and evidence for the transition to the quantum chaotic regime in 1D atomic Bloch oscillations}

Here, we provide theoretical background to the observed transition from a regular decay and revival to an irreversible decay of Bloch oscillations in the experiment. First, we discuss the statistical analysis of the energy spectrum of the tilted BH Hamiltonian \cite{KOLOVSKY03bMAT} in the parameter regime of Fig.~3 of the main article. Second, we provide numerical simulations of Bloch oscillations within the BH model that can be directly compared to the experiment.

\subsection{Spectral analysis}

\begin{figure}
\includegraphics[width=0.48\columnwidth]{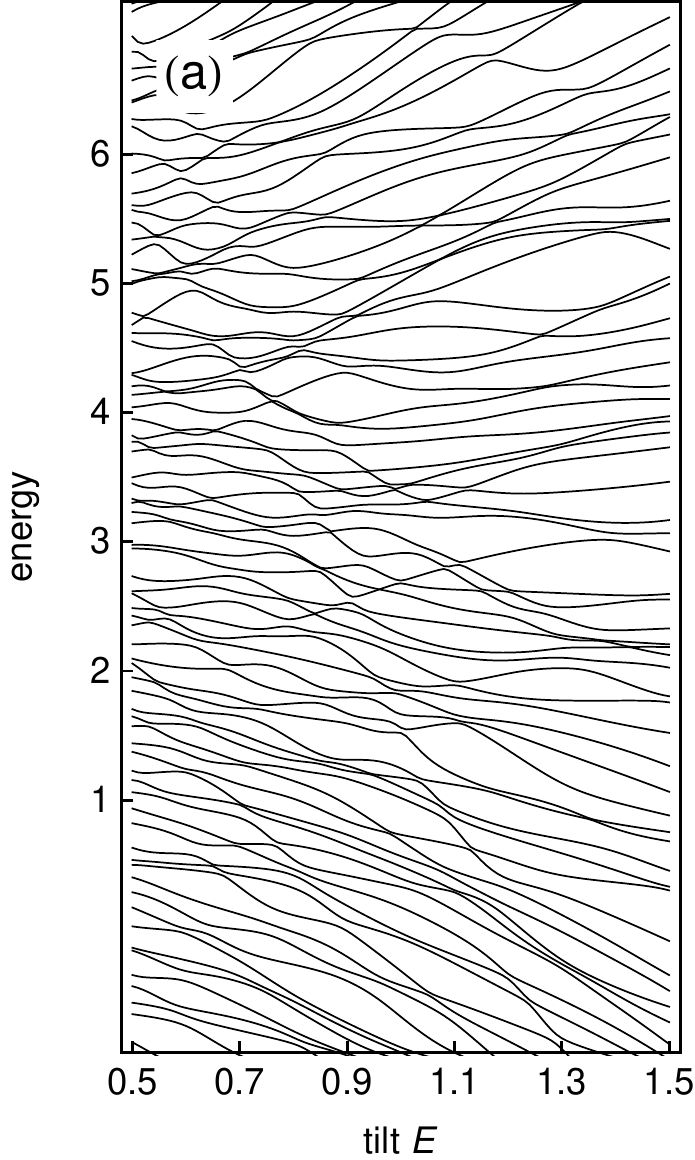}
\hspace{2mm}
\includegraphics[width=0.48\columnwidth]{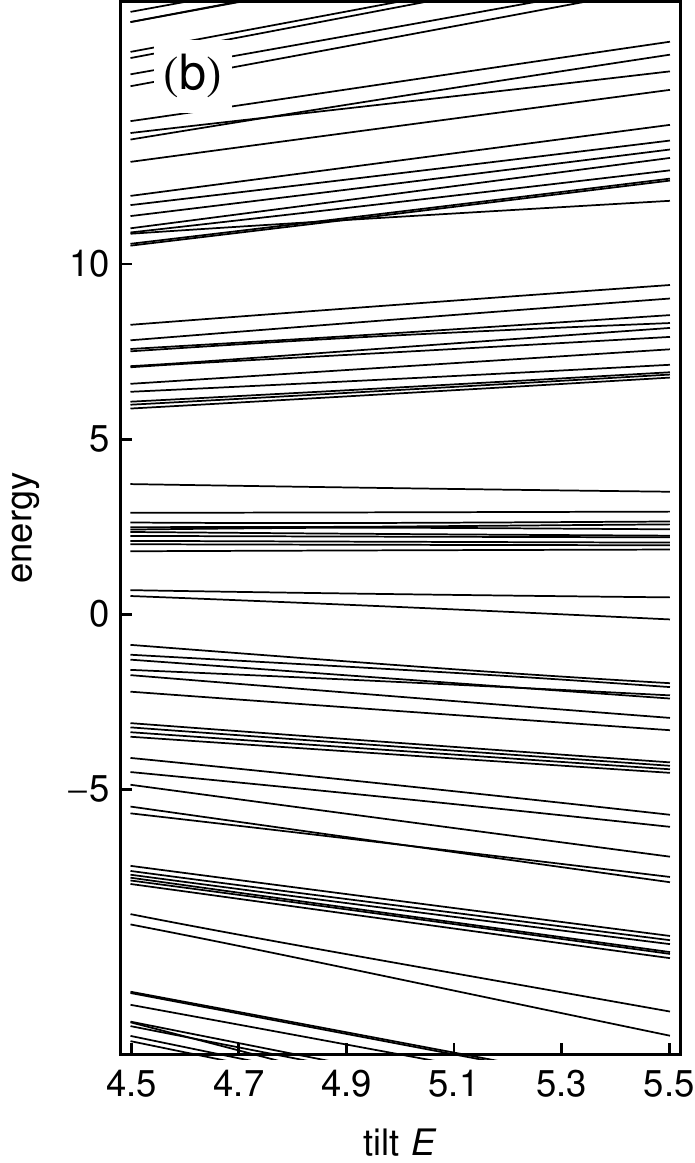}
\caption{\label{FIGAPP1} Central part of the energy spectrum of the tilted BH Hamiltonian as a function of $E$ with $J=U=1$ and $N=L=5$. The regime of small tilt (a) is contrasted to the situation of large tilt (b).}
\end{figure}

Following the approach presented in Ref.~\cite{KOLOVSKY03bMAT}, we evaluate the energy spectrum of the tilted BH Hamiltonian, Eq.~1 of the main article, as a function of the applied tilt $E$. We set $J=U=1$ in the calculation, which approximates the experimental situation of Fig.~3. In Fig.~\ref{FIGAPP1}(a) and (b) the central part of the energy spectrum is plotted as a function of $E$ (given in units of $J$) in the regime of small and large tilt for a system of $N=5$ atoms in $L=5$ lattice sites. One clearly identifies the strong non-perturbative mixing of energy levels when all energy scales $J$, $U$, and $E$ are of comparable magnitude. For a fixed value of $E$ we now compute the cumulative nearest neighbor energy level spacing
\begin{equation}
I(s) = \int_{0}^{s} P(\tilde{s}) d\tilde{s}
\end{equation}
where $P(\tilde{s})$ is the probability distribution of the normalized nearest neighbor energy level intervals $\Delta E/$ $\overline{\Delta E}$. The result for $I(s)$ for three different values of $E$, calculated in the central part of the spectrum, is shown in Fig.~\ref{FIGAPP2}(a) for a system of $N=L=7$. The numerical data is compared to Poissonian and Wigner-Dyson statistics for a Gaussian orthogonal ensemble (GOE), which are given by $P_{\rm{P}}(s)=\exp(-s)$ and $P_{\rm{GOE}}(s)=\frac{\pi}{2}  s  \exp(- \pi s^2/4)$, respectively. We identify the change from Poissonian to GOE level statistics with decreasing $E$, an indication for the onset of quantum chaos \cite{KOLOVSKY03bMAT,GiannoniBookMAT}.\\

Next, we characterize the crossover from regular to chaotic level statistics in more detail by evaluating the deviation of the numerically computed $I(s)$ from Poissonian and GOE statistics $\sigma_{\rm{P,GOE}} = 1/ \tilde{N} \int_{0}^{\infty} \left( I(s) - I_{\rm{P,GOE}}(s) \right) ^2 ds$. The normalization constant is $\tilde{N}=\int_{0}^{\infty} \left( I_{\rm{P}} - I_{\rm{GOE}}\right) ^2 ds$. The result for three different mean site occupation numbers $n=0.75$, $1$, and $1.33$ is shown in Figs.~\ref{FIGAPP2}(b)-(d), respectively. The numerical data for $\sigma$ clearly reflect the crossover from Poissonian to GOE level statistics with decreasing $E$. We notice a systematic shift of the crossover to larger values of $E$ with increasing $n$. The transition to irreversibly decaying Bloch oscillations, as observed in the experiment, is found in good agreement with the onset of chaotic spectral statistics for $n \gtrsim 1$.

\subsection{Numerical simulations of Bloch oscillation dynamics}

\begin{figure}
\includegraphics[width=1\columnwidth]{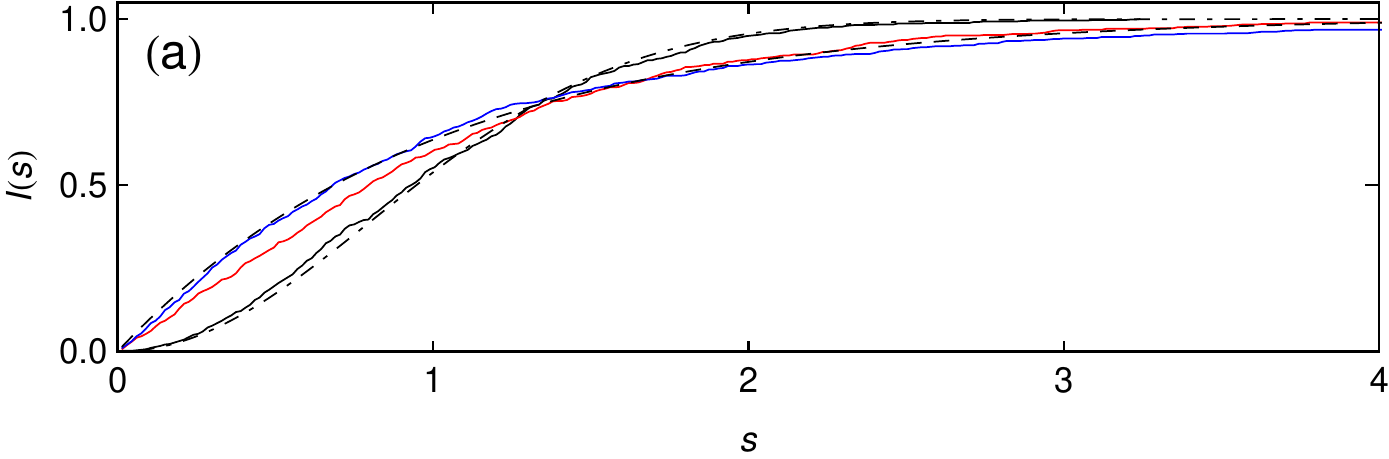}\\
\vspace{2mm}
\includegraphics[width=0.32\columnwidth]{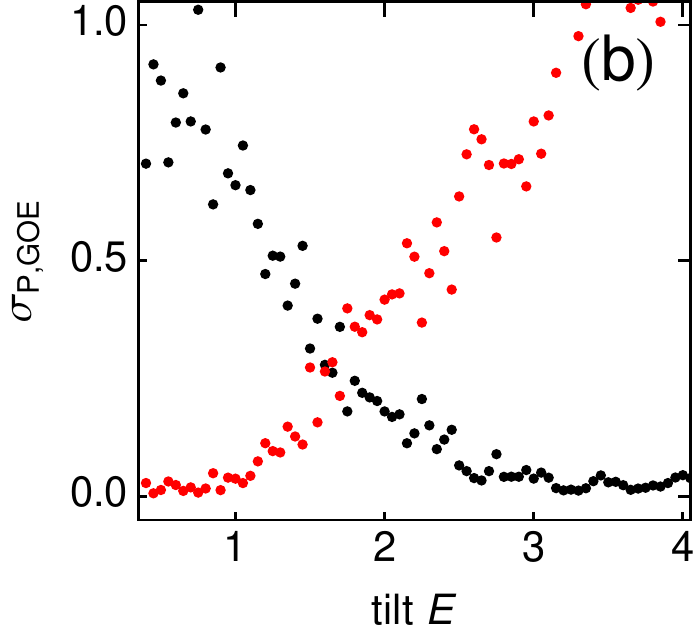}
\includegraphics[width=0.32\columnwidth]{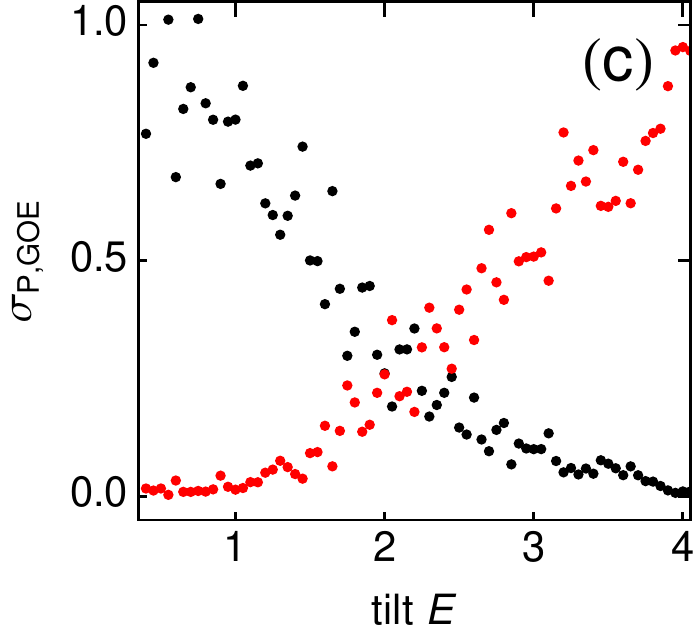}
\includegraphics[width=0.32\columnwidth]{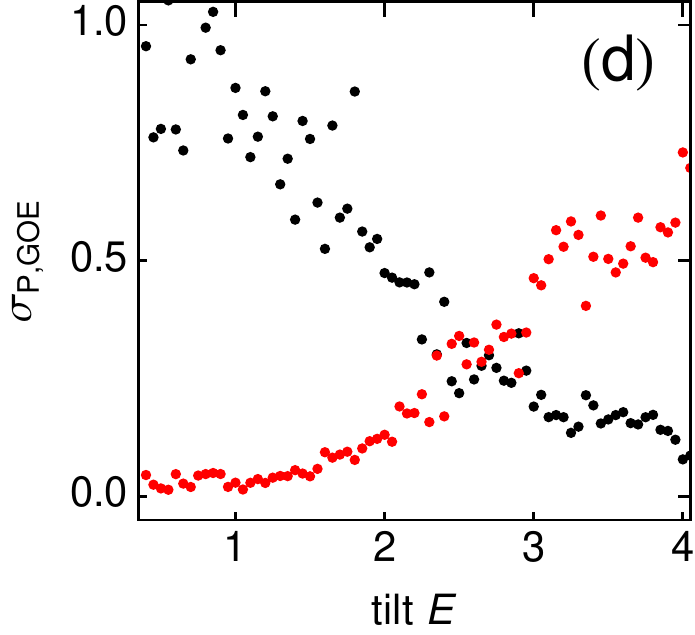}
\caption{\label{FIGAPP2} Statistical analysis of the energy spectrum. (a) Cummulative nearest neighbor level spacing distribution $I(s)$ for $E=1$ (black), $3$ (red), and $5$ (blue) with $U=J=1$ for a system of $N=7$ atoms on $L=7$ lattice sites. The dashed and dash-dotted lines show Poisson and GOE cummulative distributions, respectively. (b-d) Crossover from regular to chaotic level statistics when decreasing the tilt $E$ for fixed $U=J=1$. The deviation $\sigma_{\rm{P,GOE}}$ of $I(s)$ from Poissonian (black) and GOE (red) cummulative level spacing distribution is plotted as a function of $E$ for a system with $N=6$, $L=8$ (b), $N=L=7$ (c), and $N=8$, $L=6$ (d).}
\end{figure}

We simulate Bloch oscillation dynamics via exact diagonalization of the finite size tilted BH system for our experimental parameters in analogy to Ref.~\cite{privatMAT}. For $E \gg J,U$ tunneling is  suppressed and thus the time evolution of the initial many-body state is essentially driven by on-site dynamics. This changes when the tilt is decreased to $E \approx J, U$. Tunneling then causes boundary effects in the simulation arising from the finite size of the system. Therefore, we first calculate the ground state for a non-tilted lattice with $N$ atoms on $L$ sites for a certain combination $J$ and $U$, which we place into a larger lattice with $M>L$ sites. This constitutes the initial condition at $t=0$. We then compute the time evolution with a superimposed tilt $E$. From the many-body wavefunction $|\Psi(t) \rangle$ the mean atomic momentum
\begin{equation}
\frac{\langle p \rangle (t)}{\hbar k} = \frac{m}{N \hbar k} \frac{4 J d}{i \hbar} \, \langle \Psi(t) | \sum \limits_{j=1}^{M-1}  \hat{a}_{j+1}^{\dagger} \hat{a}_{j} + h.c.  |\Psi(t) \rangle
\end{equation}
is derived. In Fig.~\ref{FIGAPP3} we plot the computed $\langle p \rangle$ as a function of time $t$ for different $E$ and compare it to the experimental datasets that are analyzed to produce the data shown in Fig.~3 of the main article. For $E \gg U,J$ the simulation data exhibit Bloch oscillations with pronounced decay and revival dynamics. With decreasing $E$ the amplitude of the revival decreases and finally vanishes, resulting in an irreversible decay of the Bloch oscillations. Additional small oscillations can be attributed to the finite size of the sample. Note that the crossover regime in the Bloch dynamics is accompanied by the onset of chaotic level statistics in Fig.~\ref{FIGAPP2}(c).

\begin{figure}
\includegraphics[width=1\columnwidth]{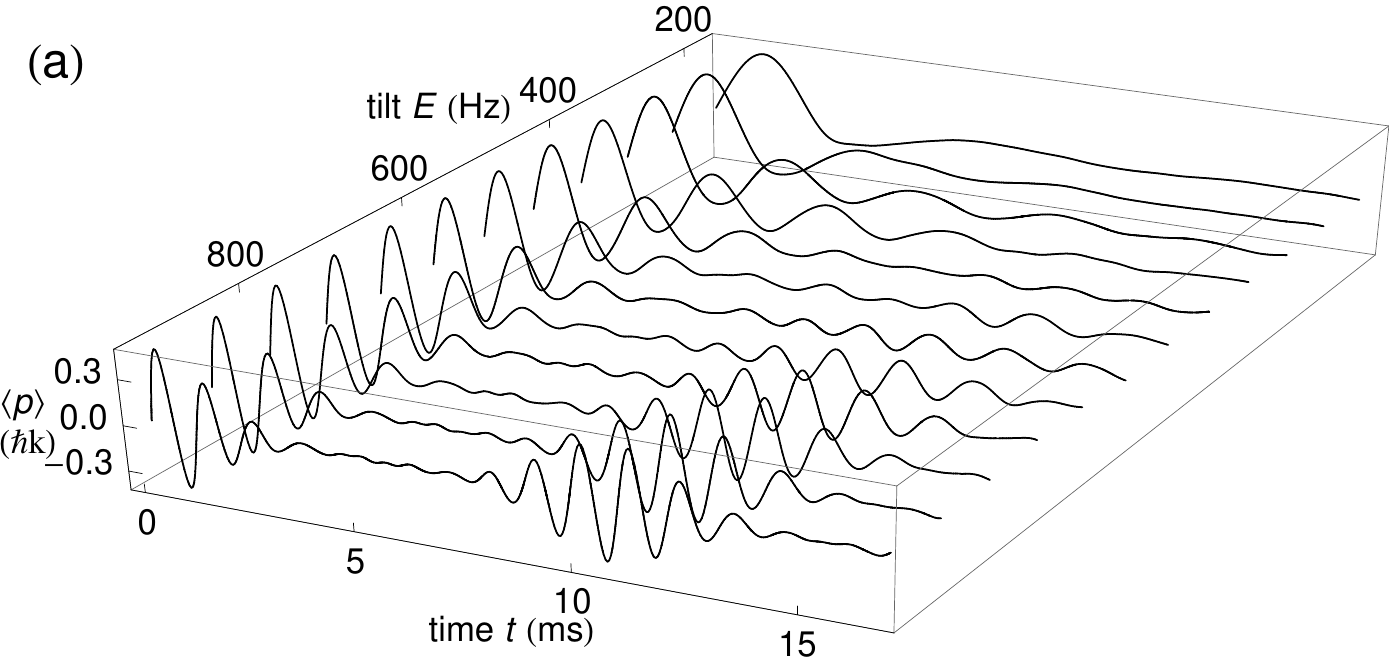}\\
\vspace{2mm}
\includegraphics[width=1\columnwidth]{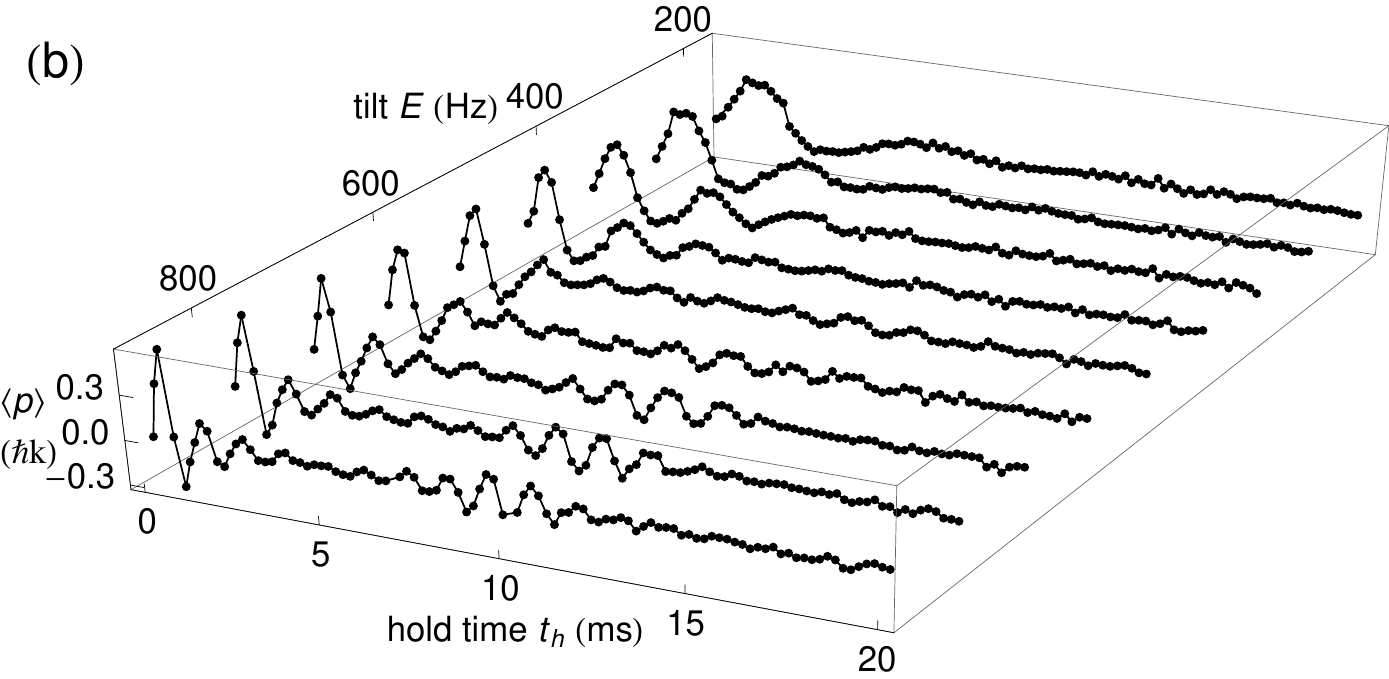}
\caption{\label{FIGAPP3} Transition from a regular decay and revival to an irreversible decay of Bloch oscillation in the experiment (b) and numerical simulation (a). The experimental datasets in (b) are the ones used for Fig.~3 of the main article. The simulations in (a) show results for a system of initially $N=7$ atoms in $L=7$ lattice sites that is propagated in time in a lattice with $M=10$ sites. Here, $U/J=0.93$, which corresponds to the experimental parameters.}
\end{figure}

The generic behavior in the experimental data is closely reproduced by our numerical simulation. The main difference is the lack of the overall damping due to the weak harmonic confinement. Including this effect in the calculation extends the simulation to experimental system sizes, which requires much more elaborate numerical methods. For the sake of a quantitative comparison we model the trap by multiplying the numerical data with the single-particle damping $\exp(-t^2/(2 \tau_0^2))$ discussed in the main text. From that we proceed in analogy to the experimental data analysis and extract a characteristic exponential dacay time $\tau$ of the initial dacay in $\langle p \rangle (t)$ and the revival width $\delta p$. The shaded areas in Fig.~3(c) of the main text reflect numerical results with varying mean site occupation $1 \leq n \leq 1.4$. Numerical data for $E \lesssim 300$ Hz are excluded from this analysis as the assignment of an exponential damping rate becomes delicate when the Bloch period exceeds $\tau$.

In Fig.~4(b) of the main article, we provide a numerical prediction of $\gamma = 1/\tau$ that is based on simulations of Bloch oscillation dynamics now with varying $U$ at fixed $E/J=3.03$ in a system with $N=L=7$ and $M=9$ \footnote{Note that for this value of $E$ it appears sufficient to enlarge the initial lattice by one site in both directions to suppress boundary effects.}. Fitting a proper decay rate to the simulated dynamics is hindered by residual small oscillations due to the finite system size especially for large values of $U$ for which strong damping ensues. Alternatively, we fit the first minimum in $p(t)$ at $t_{\rm{m}}$ and evaluate $\gamma_{\rm{m}} = -\ln \left( p(t_{\rm{m}}) / p_{U=0}(t_{\rm{m}})  \right) 1/ t_{\rm{m}}$ \cite{privatMAT}.

\bibliographystyle{apsrev}

\begin{references}
\bibitem{units} We give all energies in frequency units.
\bibitem{init} The analytic expression is derived approximating the initial state with a product of Bloch waves with quasi momentum $k=0$.
\bibitem{width} We evaluate $\delta p = \sqrt{\sum_i ( \langle p \rangle_i - \overline{\langle p \rangle} )^2/(N-1)}$ where the summand is over the sample of $N$ datapoints in the range $8.8 \, \rm{ms} \leq t \leq 10.8 \, \rm{ms}$, and $\overline{\langle p \rangle}$ denotes the mean over the sample.
\end{references}

\clearpage

\end{document}